\documentclass[a4paper]{jpconf}
\usepackage{graphicx,units,cite,booktabs,multirow,array}
\usepackage[colorlinks,allcolors=black]{hyperref}

\newcommand{\tsub}[1]{\textnormal{\tiny{#1}}}
\newcommand{\statels}[3]{\mbox{${}^{#1}$#2${}_{#3}$}}

\begin{document}

\title{Evaluation of trap-induced systematic frequency shifts for a multi-ion optical clock at the $10^{-19}$ level }

\author{J~Keller, T~Burgermeister, D~Kalincev, J~Kiethe and~T~E~Mehlst\"aubler}
\address{Physikalisch-Technische Bundesanstalt, Bundesallee 100, D-38116 Braunschweig, Germany}

\ead{tanja.mehlstaeubler@ptb.de}

\begin{abstract}
  In order to improve the short-term stability of trapped-ion optical clocks, we are developing a frequency standard based on ${}^{115}$In${}^+$ / ${}^{172}$Yb${}^+$ Coulomb crystals. For this purpose, we have developed scalable segmented Paul traps which allow a high level of control for multiple ion ensembles. In this article, we detail on our recent results regarding the reduction of the leading sources of frequency uncertainty introduced by the ion trap: 2nd-order Doppler shifts due to micromotion and the heating of secular motion, as well as the black-body radiation shift due to warming of the trap. We show that the fractional frequency uncertainty due to each of these effects can be reduced to well below $10^{-19}$.
\end{abstract}

\section{Introduction}
Laser-cooled, trapped ions are among the most successful references for optical frequency standards, achieving fractional frequency uncertainties below $10^{-17}$ \cite{Chou2010a,Huntemann2014}. However, current implementations of optical ion clocks are based on a single reference ion, which results in an intrinsically low signal-to-noise ratio. Averaging times of days to weeks are necessary for resolving the transition frequency well enough to benefit from these low systematic uncertainties. This impediment to applications can be overcome by interrogating an ensemble of ions simultaneously. To this end, we are developing a multi-ion clock based on ensembles of linear Coulomb crystals in a segmented Paul trap \cite{Herschbach2012}. Our approach is suitable for any configuration in which the electric quadrupole shift from the field gradients within the crystals can be controlled \cite{Arnold2015} or is intrinsically low, such as in the \statels{1}{S}{0}~$\leftrightarrow$~\statels{3}{P}{0} lines in group 13 ions (e.g. In${}^+$, Al${}^+$), as well as the \statels{2}{S}{1/2}~$\leftrightarrow$~\statels{2}{F}{7/2} E3 transition in Yb${}^+$.

\begin{figure}
\centerline{\includegraphics[width=.6\textwidth]{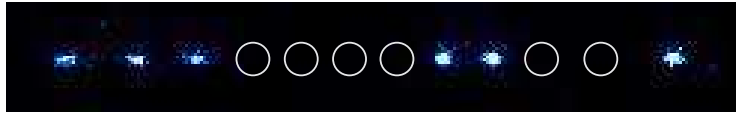}}
\caption{\label{mixedspeciesphoto}Example of a mixed-species Coulomb crystal in the prototype trap, consisting of 6 ${}^{115}$In${}^+$ ions (dark, positions indicated by circles) and 6 fluorescing ${}^{172}$Yb${}^+$ ions (from \cite{Pyka2014}).}
\end{figure}

The challenge to maintain low systematic frequency uncertainties in such a clock based on a strongly coupled many-body system is addressed in two ways: The development of scalable ion traps with low heating rates, low on-axis micromotion and a well-known thermal environment, and experiments to gain an improved understanding of the Coulomb crystal dynamics. In the current implementation, we use ${}^{115}$In${}^+$ ions, sympathetically cooled with ${}^{172}$Yb${}^+$ (see Fig.~\ref{mixedspeciesphoto}). As detailed in \cite{Herschbach2012}, indium is chosen for its high mass, low sensitivity to electric \cite{Safronova2011} and magnetic \cite{Becker2001,Herschbach2012} fields, and the facilitated scalability through direct state detection and second-stage cooling on the \statels{1}{S}{0}~$\leftrightarrow$~\statels{3}{P}{1} transition.

In this article, we present results concerning the reduction of three sources of uncertainty to well below $1\times10^{-18}$, which are also among the limiting effects in current single-ion clocks. Section \ref{EMM_section} discusses the 2nd-order Doppler shift due to excess micromotion. In section \ref{heating_rate_section}, we present a measurement of external heating rates in a prototype trap \cite{Pyka2014}. The AC Stark shift due to black-body radiation is discussed in section \ref{BBR_section}, and the results are summarized in section \ref{summary}.

\section{\label{EMM_section}Excess micromotion}
Micromotion consists of fast oscillations driven by the rf field in a Paul trap when the ion position deviates from the electric field node. It can affect the clock transition frequency through both a 2nd-order Doppler and Stark shift, the latter of which is negligible in In${}^+$ and Al${}^+$ due to the low differential static polarizability of the clock transitions \cite{Safronova2011}. Precise knowledge of rf field amplitudes at the ion positions is therefore required, in particular for the storage of extended crystals. We investigated three techniques for micromotion determination in terms of their resolution \cite{Keller2015}. In this section, we describe the two central results of this study: the explanation of an offset at a level of $10^{-19}$ seen in resolved-sideband measurements at Doppler temperature and the experimental verification of a newly derived model for the quantitative application of the photon-correlation technique in the commonly encountered regime of comparable linewidth and drive frequency. The (wrongful) application of the only previously published model \cite{Berkeland1998} under these conditions would lead to an underestimation of the micromotion-induced frequency shift by an order of magnitude and a 1st-order sensitivity to laser detuning. With this new understanding, we were able to show that both techniques allow uncertainties in the 2nd-order Doppler shift due to excess micromotion (EMM) well below $10^{-19}$.

\subsection{Resolved sideband method}
\begin{figure}
  \centerline{\textbf{a)}\includegraphics[width=.35\textwidth]{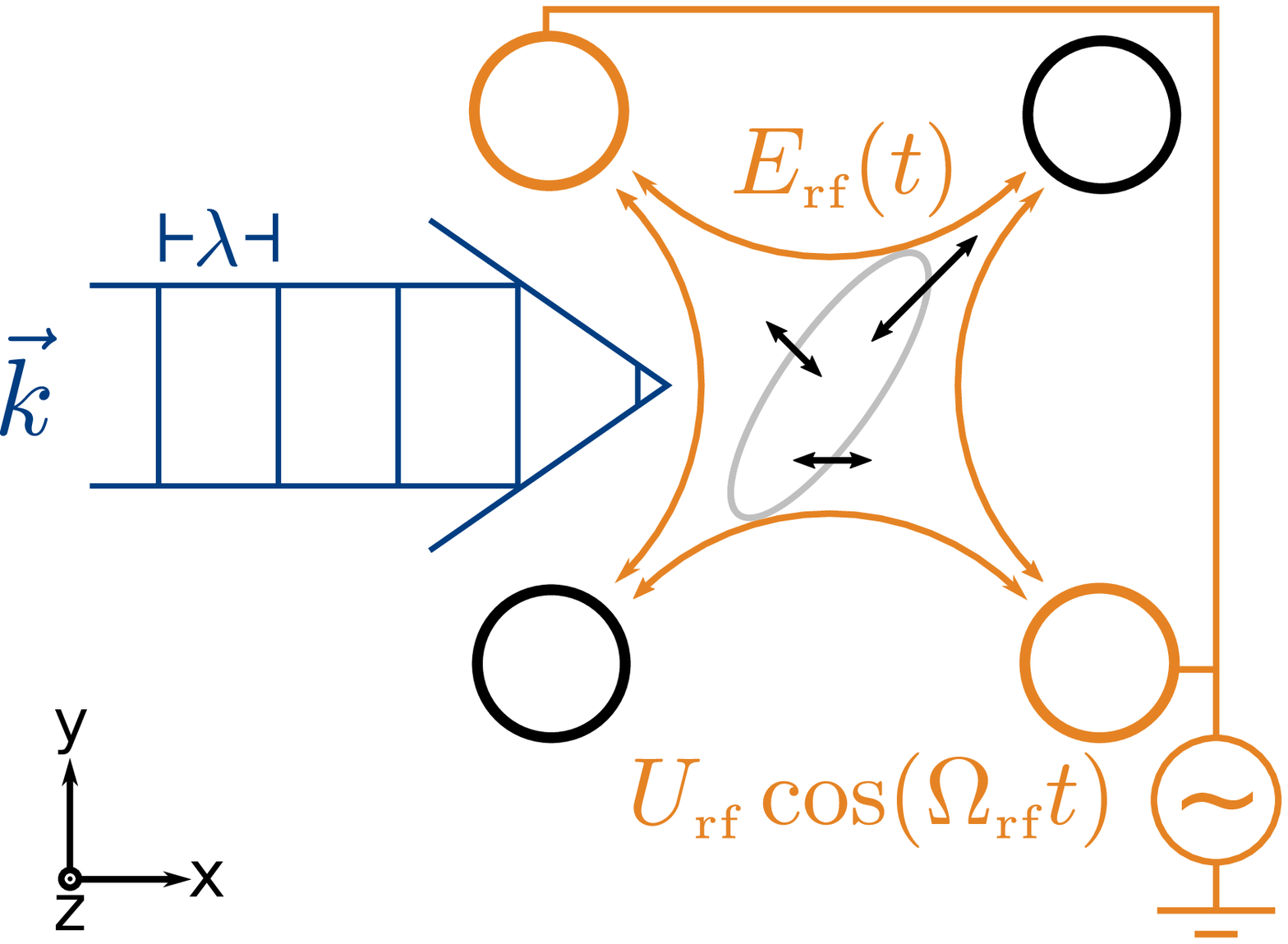}\hspace{.025\textwidth}\textbf{b)}\includegraphics[width=.445\textwidth]{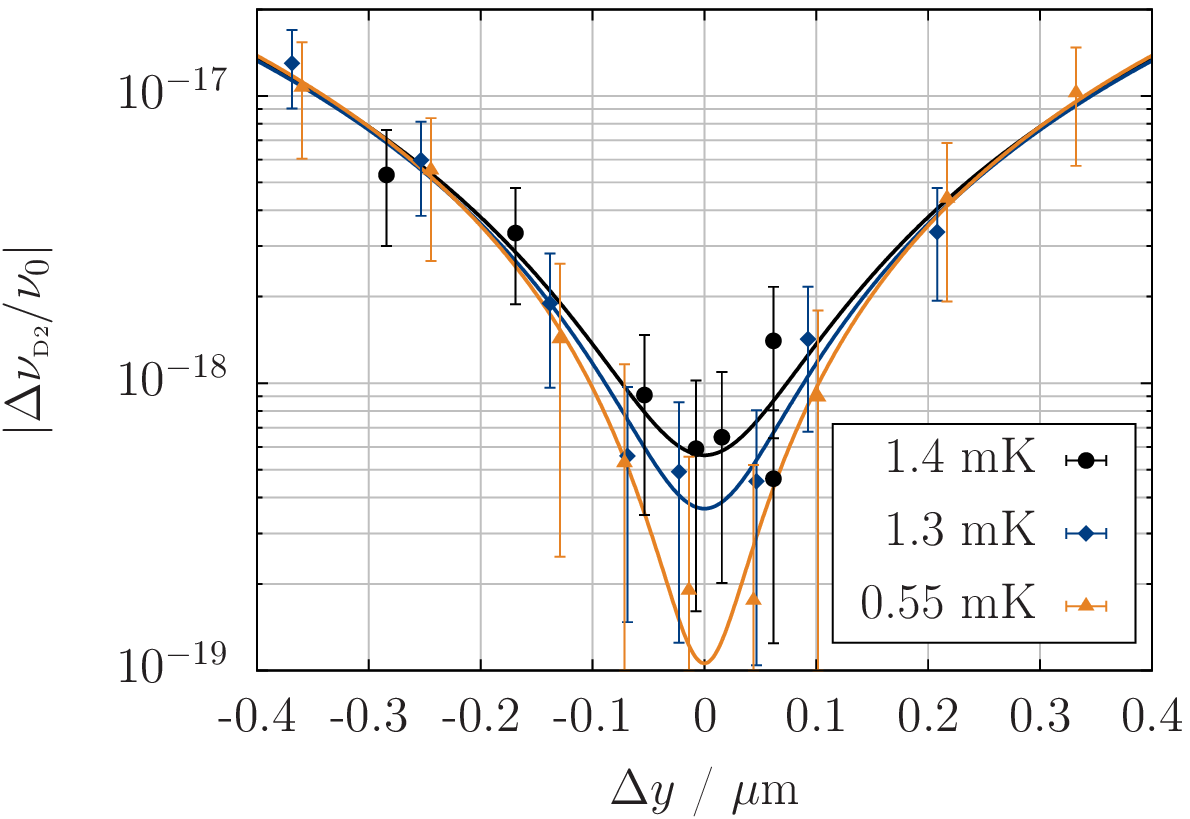}}
  \caption{\label{micromotion_IMM_sideband}Sampling of intrinsic micromotion in resolved sideband measurements. \textbf{a)} Along its secular motion trajectory (grey line), the ion undergoes intrinsic micromotion (black arrows). The correlated components of the two along $\vec{k}$ produce a temperature-dependent sideband at $\Omega_\tsub{rf}$. \textbf{b)} Experimental observation of the offset at a secular frequency of $\omega_\tsub{sec}\approx 2\pi\times\unit[500]{kHz}$ for different temperatures on the \statels{2}{S}{1/2}~$\leftrightarrow$~\statels{2}{D}{5/2} transition in a single ${}^{172}$Yb${}^+$ ion. No sideband excitation is observed at the minimum when the ion is cooled to the motional ground state.}
\end{figure}

The resolved sideband method uses a transition which is narrow compared to the sideband spectrum produced by the 1st-order Doppler shift due to micromotion. By comparison of the excitation strength on the carrier ($\Omega_0$) and micromotion sideband transitions ($\Omega_1$), a modulation index $\beta\approx2\Omega_1/\Omega_0$ is determined, which relates to the 2nd-order Doppler shift as
\begin{equation}
\frac{\Delta\nu_\tsub{D2}}{\nu_0}=-\left(\frac{\Omega_\tsub{rf}\lambda\beta}{4\pi c}\right)^2\;\textnormal{,}
\end{equation}
where $\Omega_\tsub{rf}$ denotes the trap drive frequency, $c$ the speed of light and $\lambda$ the transition wavelength. While in the low temperature limit, the evaluation of the method is straightforward and the resolution is ultimately limited by the available laser power and lowest resolvable sideband strength, the uncertainty in measurements at Doppler temperature can be limited by a temperature-dependent offset. This signal arises from the nonlinearity of $\exp(i\vec{k}(\vec{x}_\tsub{sec}+\vec{x}_\tsub{imm}))$, which mixes the phase modulation due to secular motion $x_\tsub{sec}\propto \cos(\omega_\tsub{sec}t)$ and the induced intrinsic micromotion $x_\tsub{imm}\propto \cos(\omega_\tsub{sec}t)\cos(\Omega_\tsub{rf}t)$, as illustrated in Fig.~\ref{micromotion_IMM_sideband}a. Apart from a geometrical factor $\leq1$, the expected Rabi frequency for a Fock state with $n$ phonons is
\begin{equation}
  \frac{\Omega_1}{\Omega_0}=\underbrace{\frac{q}{4}\eta^2\left(2n+1\right)+\mathcal{O}\left(\eta^4\right)}_\textnormal{intrinsic micromotion}-i\underbrace{\left\langle n\left\vert e^{ikx}\right\vert n\right\rangle\left(\frac{\beta}{2}+\mathcal{O}\left(q^3\right)\right)}_\textnormal{excess micromotion}\;\textnormal{,}
  \label{IMM_sideband_Fock}
\end{equation}
where $q$ is the Mathieu-$q$ parameter of the rf confinement (see, e.g. \cite{Major2005}) and $\eta$ is the Lamb-Dicke parameter (e.g. \cite{Wineland1998}). For a thermal state, the observed signals correspond to a thermal average of Rabi oscillations, with the minimal excitation corresponding to minimized EMM due to the $\unit[90]{{}^\circ}$ phase difference of the two terms in (\ref{IMM_sideband_Fock}). Figure~\ref{micromotion_IMM_sideband}b shows an experimental observation of the temperature dependence.

\subsection{Photon-correlation method}
\begin{figure}
  \centerline{\textbf{a)}\includegraphics[width=.445\textwidth]{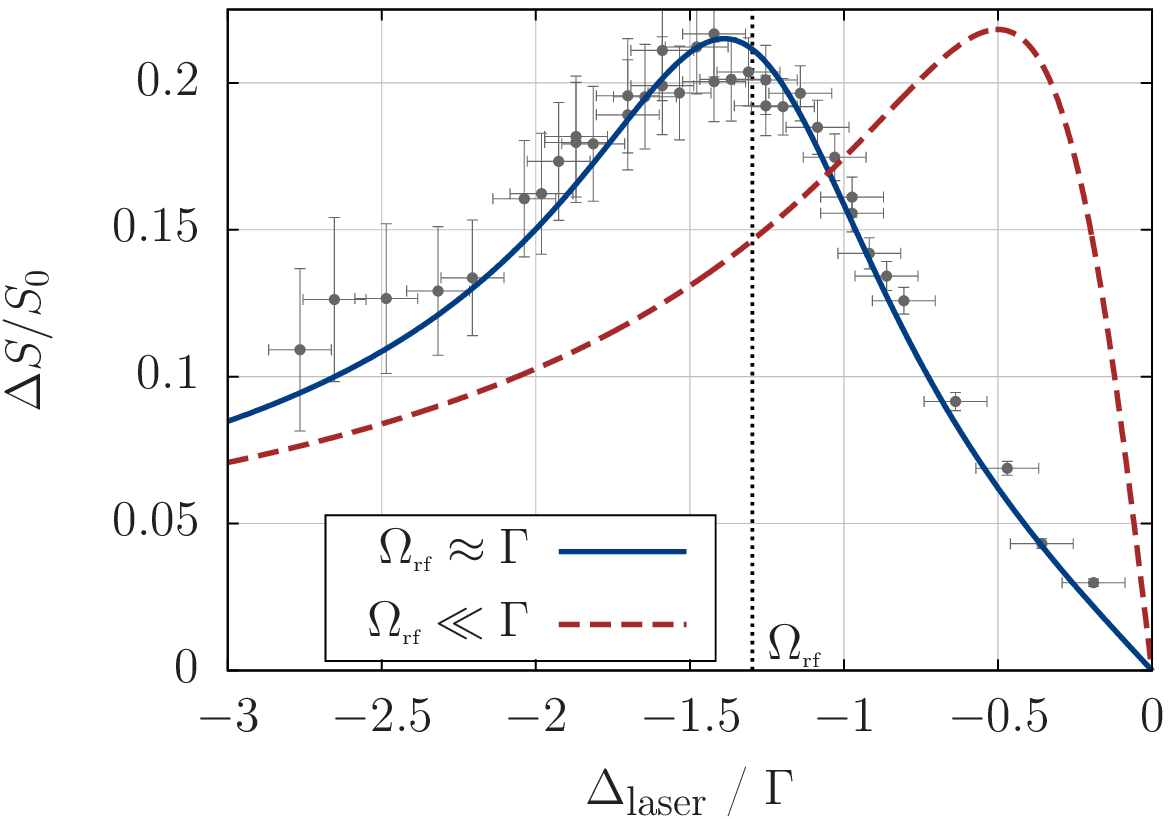}\hspace{.025\textwidth}\textbf{b)}\includegraphics[width=.475\textwidth]{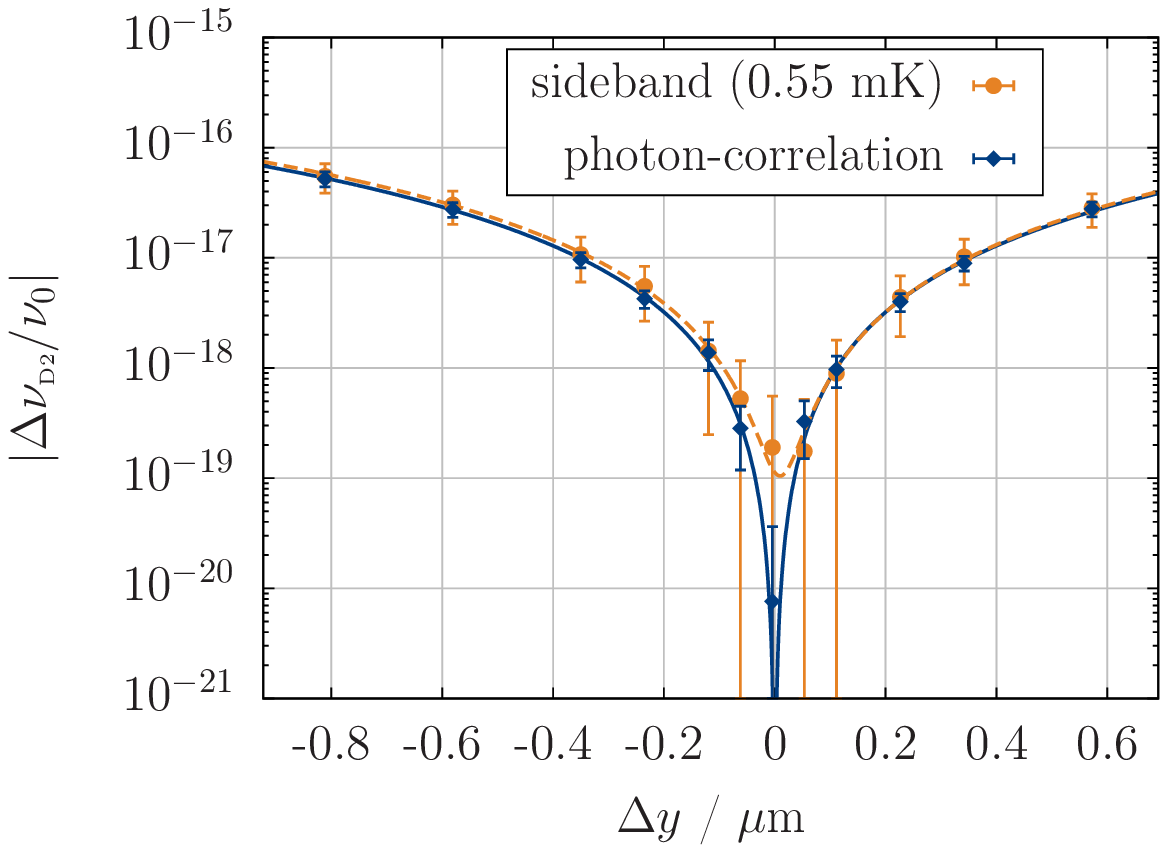}}
  \caption{\label{fmspecresultplot}\textbf{a)} Measured photon-correlation signal as a function of laser detuning (filled circles) for a fixed micromotion amplitude ($\beta=0.085$). The observed quantity is the fluorescence modulation amplitude $\Delta S$, normalized by the mean fluorescence rate $S_0$. The solid curve shows our model for the experimental parameters \mbox{$\Gamma=2\pi\times\unit[19.6]{MHz}$} and \mbox{$\Omega_\tsub{rf}=2\pi\times\unit[25.42]{MHz}$} (indicated by the vertical line). The dashed curve shows the result derived by Berkeland et al. \cite{Berkeland1998} in the limit $\Omega_\tsub{rf}\ll\Gamma$. The models agree under that condition. \textbf{b)} Quantitative comparison of photon-correlation and resolved-sideband measurements when a single ${}^{172}$Yb${}^+$ ion is radially displaced from the rf node. The residual mismatch of the fitted minima corresponds to a fractional 2nd-order Doppler shift of $3\times10^{-21}$.}
\end{figure}

The photon-correlation method employs a transition with linewidth $\Gamma$ comparable to or wider than $\Omega_\tsub{rf}$. While the evaluation had been described in the limit of $\Omega_\tsub{rf}\ll\Gamma$ by Berkeland et al. \cite{Berkeland1998}, to our knowledge no published model existed for the nowadays more common regime where $\Gamma\approx\Omega_\tsub{rf}$. Figure~\ref{fmspecresultplot} shows an experimental verification of our newly derived model for this case \cite{Keller2015} by observing the signal dependence on laser detuning (\textbf{a}) and comparison to resolved-sideband measurements when a single ${}^{172}$Yb${}^+$ ion is radially displaced from the rf node (\textbf{b}). As the method uses only the fluorescence during Doppler cooling, it is in principle applicable without pausing clock interrogation, which is advantageous in terms of servo instability \cite{Barwood2015}. Table~\ref{emmtable} summarizes the resolution limits achievable in our experimental setup for the photon-correlation and sideband methods.

\begin{table}
  \centerline{
    \scalebox{0.8}{
      \begin{tabular}{llll}
        \toprule
        method&$\Delta\nu_\tsub{D2}/\nu_0$ resolution&limited by\\\midrule
        sideband ($T_\tsub{Doppler}$)&$1.6\times10^{-19}$&sampled IMM\\\addlinespace
        sideband (ground state)&$3\times10^{-20}$&achievable Rabi frequency, decoherence\\\addlinespace
        photon-correlation&$3.5\times10^{-21}$&photon shot noise\\\bottomrule
      \end{tabular}
    }
  }
  \caption{\label{emmtable}Achieved resolution for different micromotion compensation methods.}
\end{table}

\begin{figure}
\begin{tabular}[T]{p{.4\textwidth}p{.1\textwidth}p{.5\textwidth}c}
\begin{minipage}{.5\textwidth}
\vspace{0pt}
  \centerline{\includegraphics[width=\textwidth]{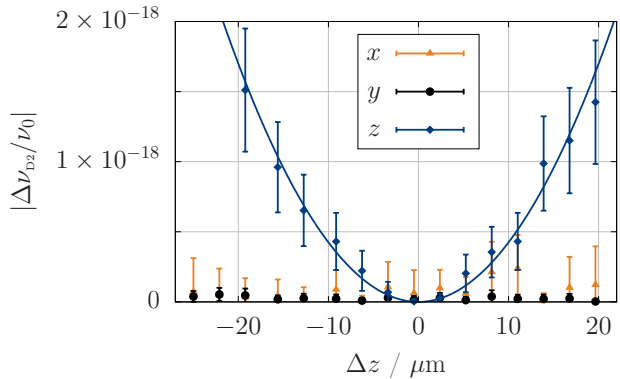}}
\end{minipage}
&
\hspace{.1\textwidth}
&
\begin{minipage}{.4\textwidth}
\vspace{0pt}
  \caption{\label{axial_emm}3D micromotion measurement as a function of the axial ion position using the photon-correlation technique. A single ion is shifted axially through a segment of the prototype trap by varying the axial dc voltages, without adjustment of the radial compensation voltages.}
\end{minipage}
\end{tabular}
\end{figure}

Since rf confinement is only used in the radial plane of a linear trap, the electric field ideally vanishes along the trap axis. In real traps, however, segmentation and machining tolerances will produce a finite axial field component, which limits the range in which clock ions can be stored without deteriorating the frequency uncertainty. If multiple ions are to be stored for spectroscopy, this field needs to be minimized by design \cite{Herschbach2012}. Figure~\ref{axial_emm} shows a 3D micromotion measurement in the prototype trap, in which a single ion was shifted along the axis, without adjustment of the radial compensation voltages.

\section{\label{heating_rate_section}Secular motion, external heating rates}
Thermally excited secular motion also contributes to the 2nd-order Doppler shift. If static contributions to the confining potential can be neglected, intrinsic micromotion doubles the resulting 2nd-order Doppler shift for the directions of rf confinement, leading to a total shift of $\Delta\nu_\tsub{D2}/\nu_0=-5k_BT/(2mc^2)$ in a linear trap. For an ${}^{115}$In${}^+$ ion sympathetically cooled to the Doppler limit of Yb${}^+$ at $\unit[0.5]{mK}$, this shift amounts to $-1.0\times10^{-18}$. A low temperature uncertainty during clock interrogation is therefore important to prevent this contribution from limiting the frequency inaccuracy. Since the different motional modes in extended crystals exhibit varying sympathetic cooling efficiency \cite{Kielpinski2000}, low external heating rates are favorable in order to ensure that the steady-state temperatures are close to the Doppler limit.

\begin{figure}
\begin{tabular}[T]{p{.45\textwidth}p{.05\textwidth}p{.5\textwidth}c}
\begin{minipage}{.5\textwidth}
\vspace{0pt}
 \centerline{\includegraphics[width=.8\textwidth]{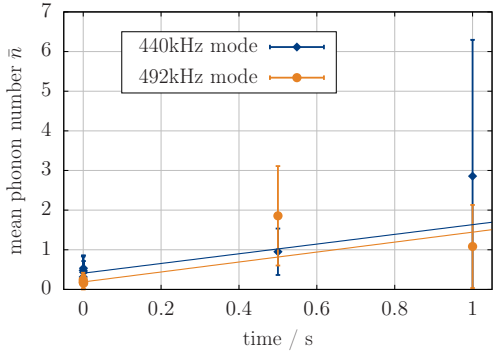}}
\end{minipage}
&
\hspace{.1\textwidth}
&
\begin{minipage}{.4\textwidth}
\vspace{0pt}
\caption{\label{heating_rates}Radial heating rate measurement in the prototype trap. The circles and diamonds show the excitation of the radial modes for a single ion as a function of time without cooling. From the fits, we determine the heating rates as $\unit[(1.2\pm0.4)]{s^{-1}}$ ($\unit[440]{kHz}$ mode) and $\unit[(1.3\pm0.5)]{s^{-1}}$ ($\unit[492]{kHz}$ mode). }
\end{minipage}
\end{tabular}
\end{figure}

External electric fields fluctuating at the secular frequency can resonantly transfer energy to the motion of the ions \cite{Brownnutt2014}. The traps used in our setup are designed to limit this effect by keeping a large distance of $\unit[\approx0.7]{mm}$ between the ions and trap electrodes and by onboard lowpass filtering of the DC voltages with a cutoff frequency of $\unit[\approx100]{Hz}$ \cite{Pyka2014}. To determine the heating rates experimentally, we cool a single ${}^{172}$Yb${}^+$ ion close to the motional ground state of both radial modes at secular frequencies of $\omega_1=2\pi\times\unit[440]{kHz}$ and $\omega_2=2\pi\times\unit[492]{kHz}$, via the \mbox{${}^2$S${}_{1/2}\leftrightarrow^2$D${}_{5/2}$} transition at $\unit[411]{nm}$. During cooling, the lifetime of the excited state is quenched using the $\unit[1650]{nm}$ \mbox{${}^2$D${}_{5/2}\leftrightarrow^2$P${}_{3/2}$} transition. After a varied amount of time, the mean phonon number in each mode is determined by comparison of 1st-order red and blue secular sidebands \cite{Turchette2000}. The result is shown in Fig.~\ref{heating_rates}: The linear fits determine similar heating rates, the higher of which is $\dot{\bar{n}}=\unit[(1.3\pm0.5)]{s^{-1}}$, corresponding to an electric field power spectral density of $S_E(\unit[492]{kHz})=\unitfrac[(1.8\pm0.7)\times10^{-14}]{(V/m)^2}{Hz}$. In the absence of cooling, the energy transferred to the ion increases the 2nd-order Doppler shift due to each mode as
\begin{equation}
  \left\vert\frac{\partial}{\partial t}\left(\frac{\Delta\nu_\tsub{D2}}{\nu_0}\right)\right\vert=\frac{\dot{\bar{n}}\hbar\omega_\tsub{sec}}{mc^2}\;\textnormal{,}
  \label{2ndorderdoppler_heatingrates}
\end{equation}
where $\omega_\tsub{sec}$ denotes the secular frequency of the respective mode and $m$ the mass of the ion. This shift amounts to \mbox{$\unit[(2.5\pm1.0)\times10^{-20}]{s^{-1}}$} and \mbox{$\unit[(1.1\pm0.4)\times10^{-19}]{s^{-1}}$} for a clock based on ${}^{115}$In${}^+$ and ${}^{27}$Al${}^+$, respectively.

\section{\label{BBR_section}Black-body radiation due to trap heating}
\begin{figure}
  \centerline{\textbf{a)}\hspace{.025\textwidth}\includegraphics[height=4cm]{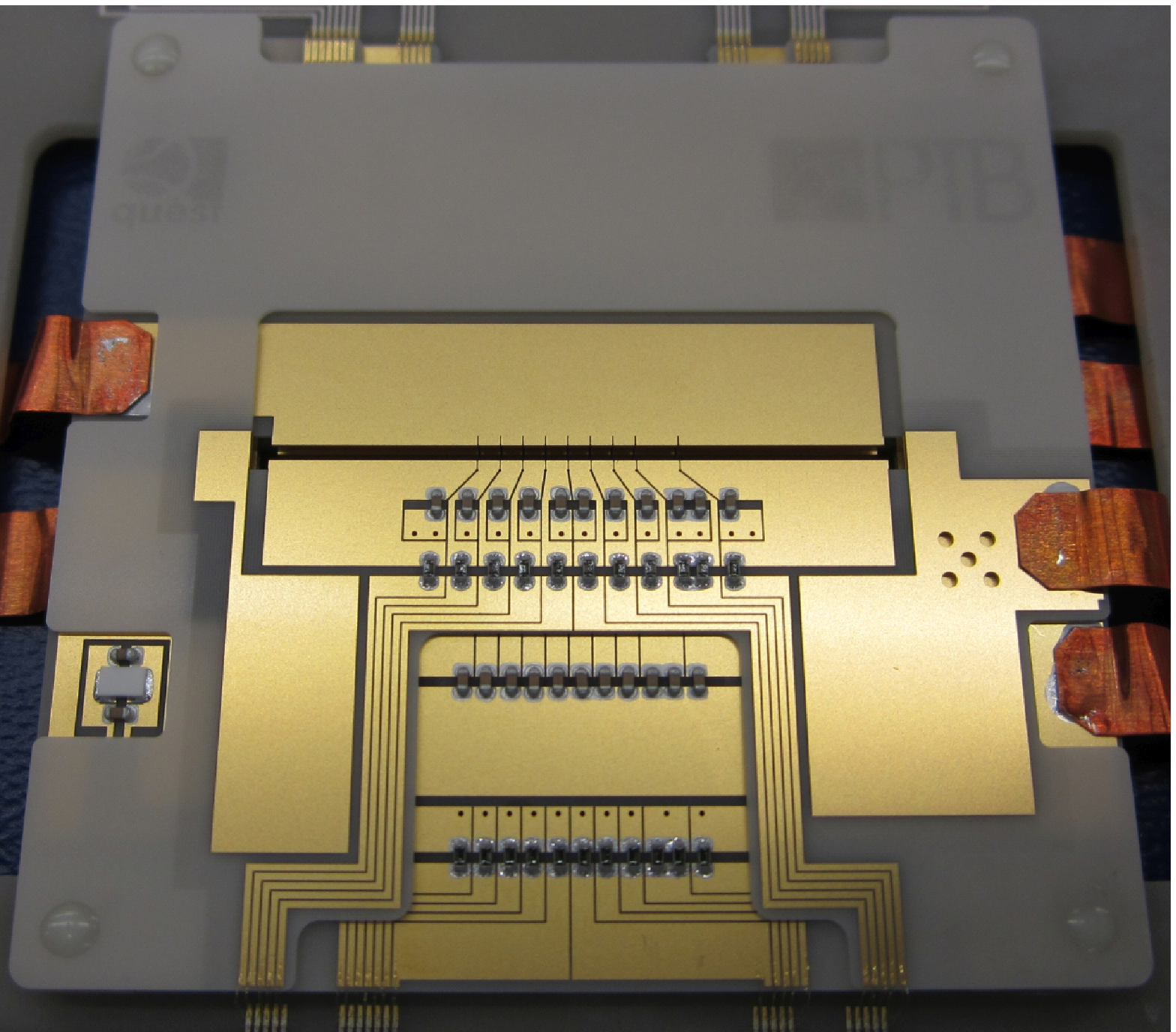}\hspace{.05\textwidth}\textbf{b)}\hspace{.025\textwidth}\includegraphics[height=4cm]{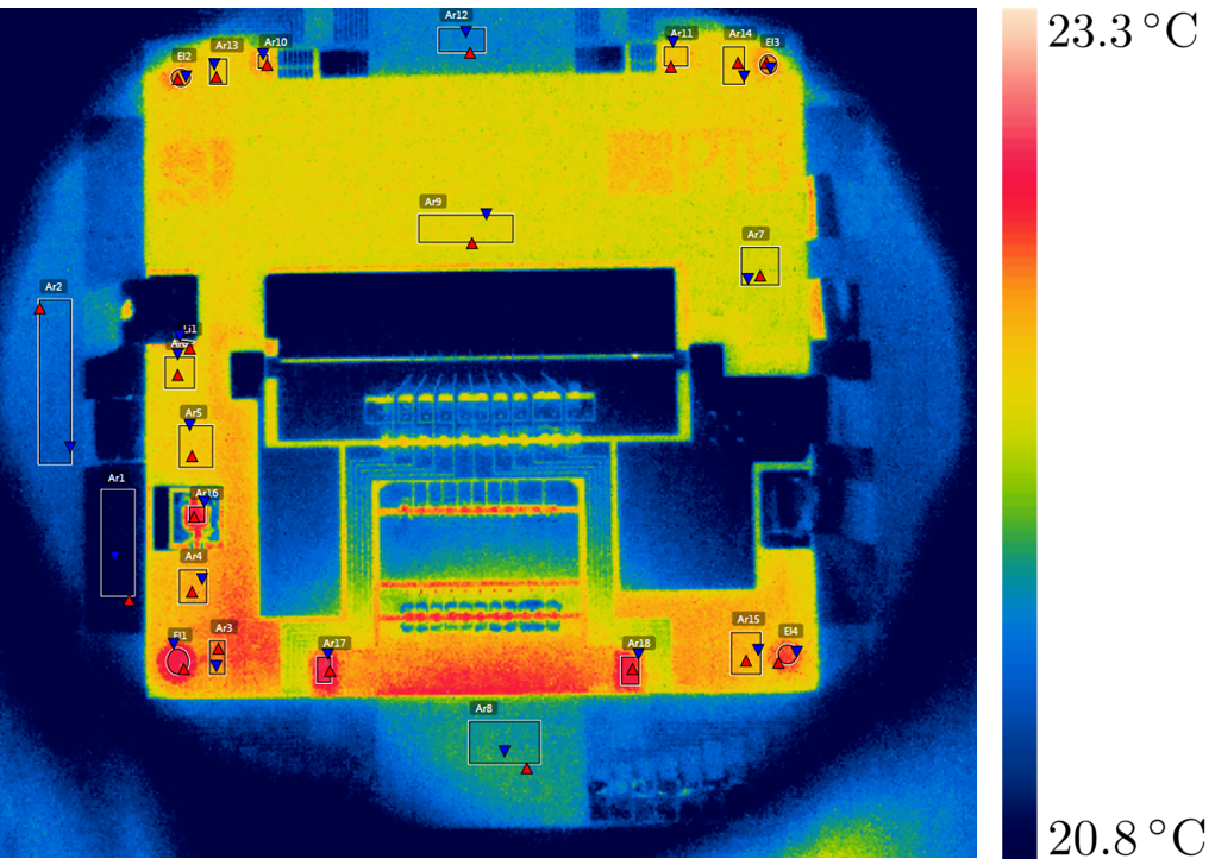}}
  \caption{\label{AlN_trap_1kV}\textbf{a)} Image of the next-generation trap made from a stack of 4 AlN wafers. Ions are trapped in the central slit. On the bottom left, one of the temperature sensors can be seen. \textbf{b)} Thermal image showing the temperature increase at $U_\tsub{rf}=\unit[1]{kV}$,  $\Omega_\tsub{rf}=2\pi\times\unit[15.4]{MHz}$.}
\end{figure}

The AC Stark shift due to black-body radiation (BBR) amounts to $1.36\times10^{-17}$ for In${}^+$ at $\unit[300]{K}$ \cite{Herschbach2012,Safronova2011}. In Paul traps, a particular concern is the warming of the trap structure due to the applied rf voltages on the order of several hundred volts. In order to reduce the corresponding uncertainty, tests have been performed on a new trap design based on AlN wafers, shown in Fig.~\ref{AlN_trap_1kV}a. Besides the use of materials with a high heat conductivity and low dielectric losses to limit the temperature increase in operation, two onboard Pt100 sensors are included for in-situ temperature measurements. The effective thermal environment experienced by the ions is deduced from a finite-element model, the parameters of which are adjusted using measurements on a test trap \cite{Dolezal2015}.

Trap temperature measurements are performed with a calibrated IR camera and thermistors. Figure~\ref{AlN_trap_1kV}b shows a thermal image of the ceramics test trap when an rf amplitude of $\unit[1]{kV}$ at $\Omega_\tsub{rf}=2\pi\times\unit[15.4]{MHz}$ is applied, which would yield trap frequencies of $\omega=2\pi\times\unit[1.0]{MHz}$ ($\omega=2\pi\times\unit[1.52]{MHz}$) for ${}^{172}$Yb${}^+$ (${}^{115}$In${}^+$). FEM simulations with experimentally adjusted material parameters predict an effective temperature increase at the ion position of $\Delta T_\tsub{ion}=\unit[1.0]{K}$, which corresponds to an AC Stark shift of $\Delta\nu_\tsub{BBR,In}/\nu_0\approx 2\times10^{-19}$. Using the calibrated Pt100 sensors, the uncertainty of the temperature rise seen by the ions is reduced to $\unit[0.1]{K}$, which corresponds to a BBR shift uncertainty of $2\times10^{-20}$ \cite{Dolezal2015}. The total uncertainty of the BBR shift in In${}^+$ is then limited by the theoretical prediction of the differential polarizability \cite{Safronova2011}. This can be reduced with an experimental determination, e.g. by controlled heating of the trap via the Pt100 sensors during interrogation.

\section{\label{summary}Summary}
This article presents our recent results concerning the reduction of the major trap-related sources of frequency uncertainty in current ion clocks. Our analysis of micromotion measurement techniques has shown that the corresponding 2nd-order Doppler shift can be determined with an uncertainty well below $10^{-19}$, independent of the ion mass, with both the resolved-sideband and photon-correlation methods \cite{Keller2015} (see Table~\ref{emmtable}). The heating rates of less than $2$ phonons per second measured in a prototype trap facilitate sympathetic cooling in extended mixed-species crystals and give rise to a 2nd-order Doppler shift increase of \mbox{$\vert\partial/\partial t(\Delta\nu_\tsub{D2}/\nu_0)\vert=\unit[(2.5\pm1.0)\times10^{-20}]{s^{-1}}$} in In${}^+$ if cooling is paused during interrogation. In the next-generation AlN wafer trap, onboard temperature sensors, in combination with an experimentally verified thermal model, allow a precise knowledge of the black-body spectrum at the ion position due to trap warming. The residual temperature uncertainty corresponds to a fractional uncertainty of $2\times10^{-20}$ in the BBR AC Stark shift \cite{Dolezal2015}.



\section*{References}
\bibliographystyle{iopart-num}
\bibliography{fsm_proceedings}

\end{document}